# Heavy Ion Physics at the LHC

P. Giubellino
*INFN, Torino, Italy*

The first Pb-Pb collisions at the LHC are little more than a year away. This paper discusses some of the exciting measurements which the experiments will be able to perform in the very first run, even with modest luminosity, and gives a very short overview of some of the most interesting ones attainable with more extended runs. The dedicated Heavy-Ion experiment ALICE, but also ATLAS and CMS, experiments optimized for p-p collisions, are ready and eager to make best use of the nuclear beams in the LHC as soon as they will be available. The main specificities of the three detectors for Heavy-Ion collisions will also be briefly addressed in this paper. I will try to show that already the first results obtainable with Heavy-Ion beams at the LHC will qualify it as a *discovery machine*, capable to provide fundamental new insight to our knowledge of high-density QCD matter.

## 1. INTRODUCTION

Heavy-Ion Physics is an integral part of the baseline scientific program of the CERN Large Hadron Collider (LHC), the new accelerator which, at the moment of writing is just starting operation with the first tests of injection of proton beams. When accelerating nuclear beams, the LHC will provide unprecedented conditions for the study of high-density strongly interacting matter, and will represent a unique opportunity for Heavy-Ion Physics. The LHC, in Pb-Pb collisions, will reach a center-of-mass energy of $\sqrt{s_{NN}}$ = 5.5 TeV, almost 30 times higher than the maximal energy explored at the Relativistic Heavy Ion Collider RHIC. This is an even larger increase in center of mass energy than the factor 10 obtained in going from the CERN SPS to RHIC. The higher energy will open an entirely new frontier, making very large energy densities attainable, $\varepsilon_{Bjorken} \approx 10$ GeV/fm$^3$. The large energy density, coupled with the lifetime of the plasma, which will also grow by a factor 2 to 3, will provide "ideal" conditions for the study of the plasma, closer to the conditions simulated in lattice QCD calculations. Moreover, the $x$ and $p_T$ range which will be explored will grow enormously, respectively towards very small $x$ values $O(10^{-5})$ and towards large transverse momenta, up to over 100 GeV/c. Last but not least, the very interesting field of heavy quark production will benefit of cross sections 10 to 10,000 times larger than at RHIC. The LHC will therefore bring us a major step forward towards the ultimate goal of ultra-relativistic heavy ion physics, which is to build a coherent understanding of the properties of high-density nuclear matter. The interest of the field is reflected in the constantly growing number of scientists working on the preparation of the experiments: while the members of the ALICE collaboration continue to increase, and are now well over a thousand, a sizeable community of over a hundred physicists has developed in CMS very actively pursuing the Heavy-Ion Physics program, also including dedicated new detectors, and a smaller, yet rapidly growing, one is working in ATLAS.

The rich harvest of new results expected at the LHC has been the subject of many excellent reviews in recent years, and I will not even attempt to make summary of them, nor a complete list. The proceedings of a recent CERN Theory Institute workshop provide a rather comprehensive summary of the efforts from the theory community [1], while some general papers produced by the ALICE, ATLAS and CMS collaborations [2,3,4,5] offer an overview of the potential of the experiments.

## 2. THE LHC AS AN ION COLLIDER

The LHC is the latest of a series of accelerators of nuclear beams, which have been the instruments of relativistic Heavy Ion Physics during now over two decades. Each accelerator is about one order of magnitude higher in center-of-mass energy than the previous one. From the 1 GeV per nucleon pair of the Bevalac at Berkeley, started some 20 years ago, accelerators have increased in energy to the AGS at BNL (few GeV) to the SPS at CERN (almost 20 GeV), to the Relativistic Heavy Ion Collider at BNL (which started operation in 2000 with 130 GeV, increased the following year to 200). In just little more than 20 years heavy-ion accelerators have covered an enormous path, during which experimenters have learnt how to deal with events of unprecedented complexity, with first hundreds, than thousands of particles produced. The task has been a formidable one, both from the instrumental point of view and from the point of view of the physics analysis. The phase diagram of strongly interacting matter has been populated of measured points, and strategies have been developed to investigate the nature and dynamical evolution of the high density system generated in these interactions. The LHC is yet another big step forward, and it has required a major effort to develop detectors which would be able to extract as much information as possible from the interactions.

The LHC, designed to collide protons at a c.m.s. energy $\sqrt{s}$=14 TeV, will also accelerate ions up to the same magnetic rigidity, i.e. a c.m.s. energy per nucleon pair of 5.5 TeV for Pb-Pb, and allow the study of both symmetric systems (e.g. Pb-Pb) and asymmetric collisions, such as proton-nucleus.

In table I examples of the c.m.s. energy and expected maximum luminosity in the LHC for some collision systems are given. The typical expected yearly running times are of the order of $10^7$ s for proton-proton collisions and $10^6$ s for the heavier systems. As discussed in detail in [2], the average luminosity actually delivered to each of the experiments will depend on their number and on the effectiveness of schemes, currently under study, to modulate the instant luminosity in order to maximize its average over a length of time. Overall, each experiment should count on an average luminosity not larger than one half of the peak values indicated in the table.

Table I: Max. luminosity and c.m. energy expected for different collision systems in the LHC. The geometrical cross sections are also given.

| Collision system | $\sqrt{s_{NN}}$(TeV) | $L_0$ ($cm^{-2}s^{-1}$) | $\sigma_{geom}$ (b) |
|---|---|---|---|
| p-p | 14.0 | $10^{34}$ * | 0.07 |
| Pb-Pb | 5.5 | $10^{27}$ | 7.7 |
| p-Pb | 8.8 | $10^{29}$ | 1.9 |
| ArAr | 6.3 | $10^{29}$ | 2.7 |

At present, the plan foresees a pilot run in late 2009 with Pb-Pb at about 1/20 of the nominal luminosity, followed by 2-3 years of Pb-Pb at full luminosity, necessary to collect about 1 $nb^{-1}$ of integrated luminosity in each of the experiments, followed by one year of p-Pb and eventually a year of lighter ions, depending on the results accumulated up to that point. For a longer term future, involving upgraded detectors and further improvements of the accelerator complex, several scenarios are being studied, waiting for guidance from the first physics results.

## 3. HEAVY ION EXPERIMENTS AT THE LHC

The LHC detectors and their status are described in these same proceedings [6, 7, 8], so I limit myself here to a few general comments. Contrary to RHIC, where several specialized detectors have been constructed, at the LHC only one experiment dedicated to heavy ion Physics has been built: ALICE. The experiment has therefore been designed to cover most of the relevant observables at the same time. ALICE consists of a central detector system embedded in a very large solenoid with a field of 0.5T, a forward muon spectrometer and forward multiplicity and centrality detectors. The central detector system provides fine granularity tracking over long track length, precise determination of the impact parameter and, via the combination of several identification techniques, the identification of most of the produced particles over a wide range of momenta, a unique feature among the LHC experiments. In figure 1 is shown the momentum coverage of the various ALICE PID detectors, demonstrating the extent of the momentum coverage achieved. In addition, short-lived particles are identified through their decay topologies. For example, K and Λ decays can be efficiently reconstructed up momenta beyond 10 GeV/*c*. The very fine granularity of the tracking detectors, providing a large number of space points per track, allows particle tracking with efficiency larger then 95% down to momenta below 200 MeV/*c* even at the highest particle multiplicities foreseen at the LHC, while the very large track length measured allows ALICE to achieve an excellent momentum resolution ($\delta p/p < 5\%$ at 100 GeV/*c*). Recently, the construction of an Electromagnetic Calorimeter has been approved, covering about a third of the azimuth and providing improved energy resolution and triggering for jets. A high-resolution crystal photon spectrometer is placed at a very large radial distance from the interaction point (4.7m). The muon spectrometer sits at forward rapidities, between 2.5 and 4, and features a front absorber placed very close to the interaction point (90cm) to minimize muonic background from weak hadronic decays and uses chambers with 2-D pad readout to handle the large multiplicities.

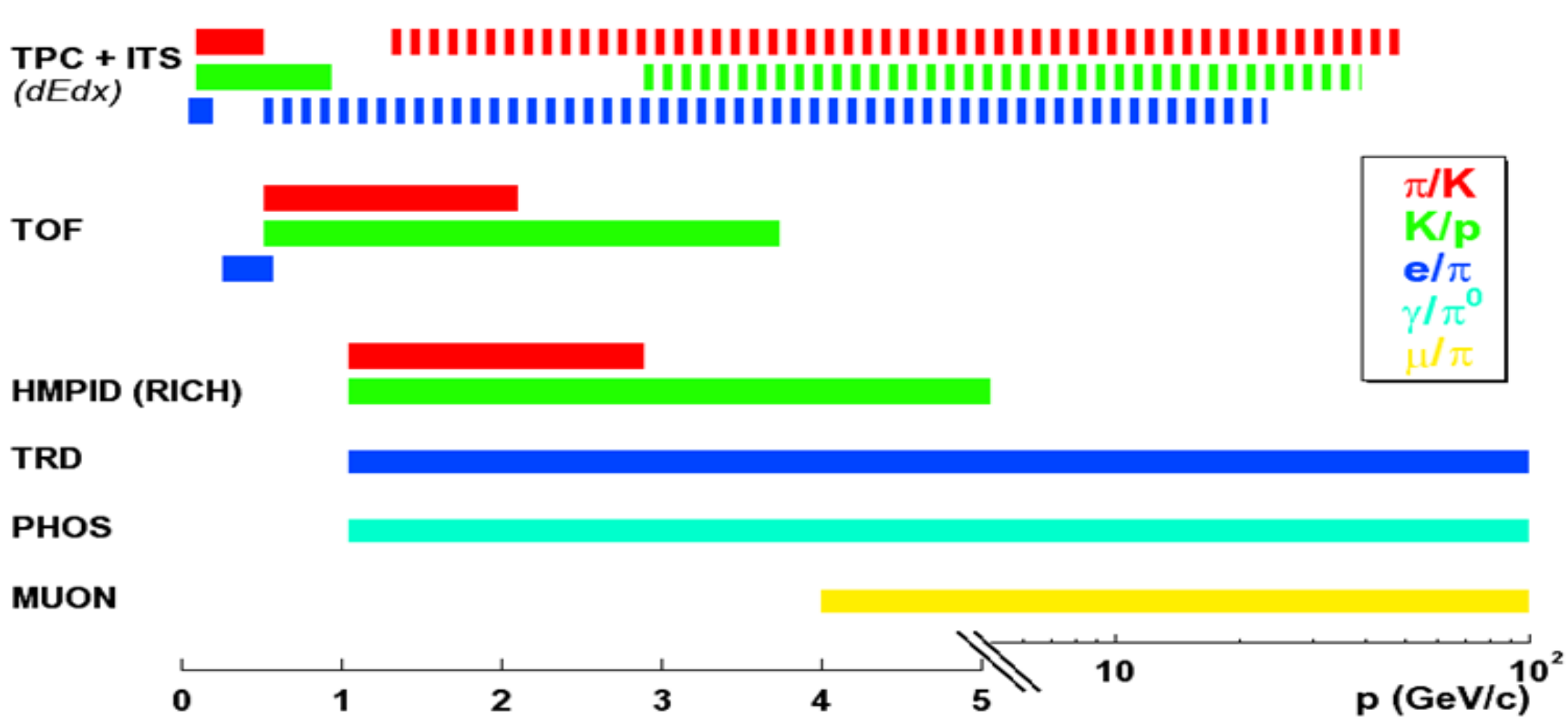

Figure 1: Particle identification in ALICE: momentum range and detectors used

ATLAS and CMS are optimized for the measurement of high-$p_t$ charged particles, including muons and electrons, high-energy photons and jets. They feature tracking in magnetic fields which are much stronger than the ALICE one, and quasi-hermetic electromagnetic and hadronic calorimetry. The momentum resolution for tracks with 100 GeV/*c* $p_t$ is about 4% for ATLAS and 2% for CMS. Being optimized for rare signals, they both have very large acceptance, more than five units of rapidity for tracking and 6 for calorimetry. These characteristics make the pp detectors largely complementary to ALICE, providing excellent data on jets (especially on di-jet and photon-jet events) and high-$p_t$ muons. CMS will also address many of the general studies (multiplicity, momentum distributions) discussed below. ALICE and CMS rely on distant (over 100 m from the interaction point) zero-degree calorimeters to determine the impact parameter of the Pb-Pb collisions.

## 4. THE FIRST "THREE MINUTES"

It is almost a tradition for ultra-relativistic heavy ion experiments to produce important results on the fly, with the very first collisions. In fact, in 1986 the experiments at the SPS produced the first spectra, demonstrating an unexpectedly high degree of stopping and large transverse energy production [9], based on an accelerator test run which took place one week *before* the official start of the heavy ion run! Equally striking has been the impact of the first collisions at RHIC: with the first run starting on June 12$^{th}$, 2000, on June 19$^{th}$ the first paper on multiplicity distributions [10] was out, excluding 90% of the existing theoretical predictions and opening the way to concepts in which the growth in particle production expected because of the semi-hard nature of the dominant processes is tamed by saturation effects. Just two month later, based on 22k events, the paper reporting the surprising large values of elliptic flow, indicating strong early-time thermalization, was published [11]. In rapid sequence, after just 3 weeks of very low luminosity running and with no previous commissioning runs with proton beams, a number of fundamental papers on RHIC results appeared. Similarly, we are confident that even a very short, low-luminosity pilot run at the LHC with Pb beams will already produce results of great scientific value. Let us take consider a few examples. First of all, the value of charged multiplicity in the central region will be measured. Already this relatively simple measurement will challenge the models, which currently describe particle production in nuclear collisions up to RHIC energies.

Figure 2, from [2], shows a collection of multiplicity results from heavy ion and p-antiproton collisions. The heavy ion data are rescaled by the number of nucleons participating in the collision, $N_{part}$. This is a phenomenological quantity (see, for example, Ref. [12]), which for pp is 2, for central AA collisions, about 2*A*, and which can be estimated as a function of impact parameter *b*. When extrapolating from existing data to LHC energies we see in the figure the striking difference between the results obtained applying a saturation model or a fit in $\ln^2\sqrt{s}$ . As discussed in detail in [14] and [15], the multiplicity measurement will by itself either force us to revisit the central ideas used to explain the particle production at RHIC or to face a violation of the universal structure observed in multiparticle production at lower energies. In either case, the very first nucleus-nucleus collisions at the LHC will already profoundly influence our understanding of the field.

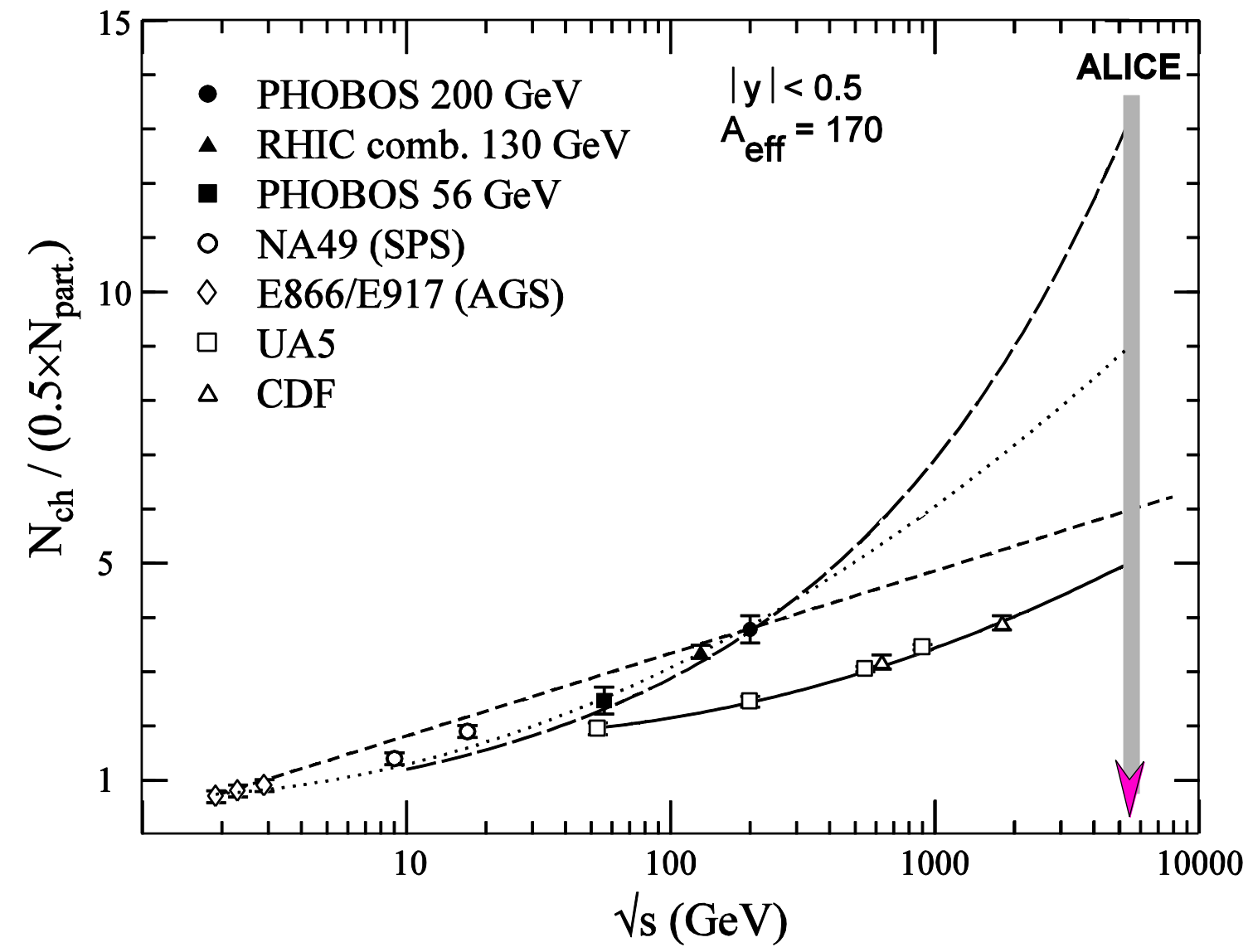

Figure 2: Charged-particle rapidity density per participant pair as a function of centre-of-mass energy for AA and pp collisions. The dashed line is a fit $0.68\ln(\sqrt{s}/0.68)$ to all the nuclear data. The dotted curve is $0.7+0.028\ln^2\sqrt{s}$. It provides a good fit to data below and including RHIC, and predicts $N_{ch}$ = 9 x 170= 1500 at LHC. The long dashed line is an extrapolation to LHC energies using a saturation model [13].

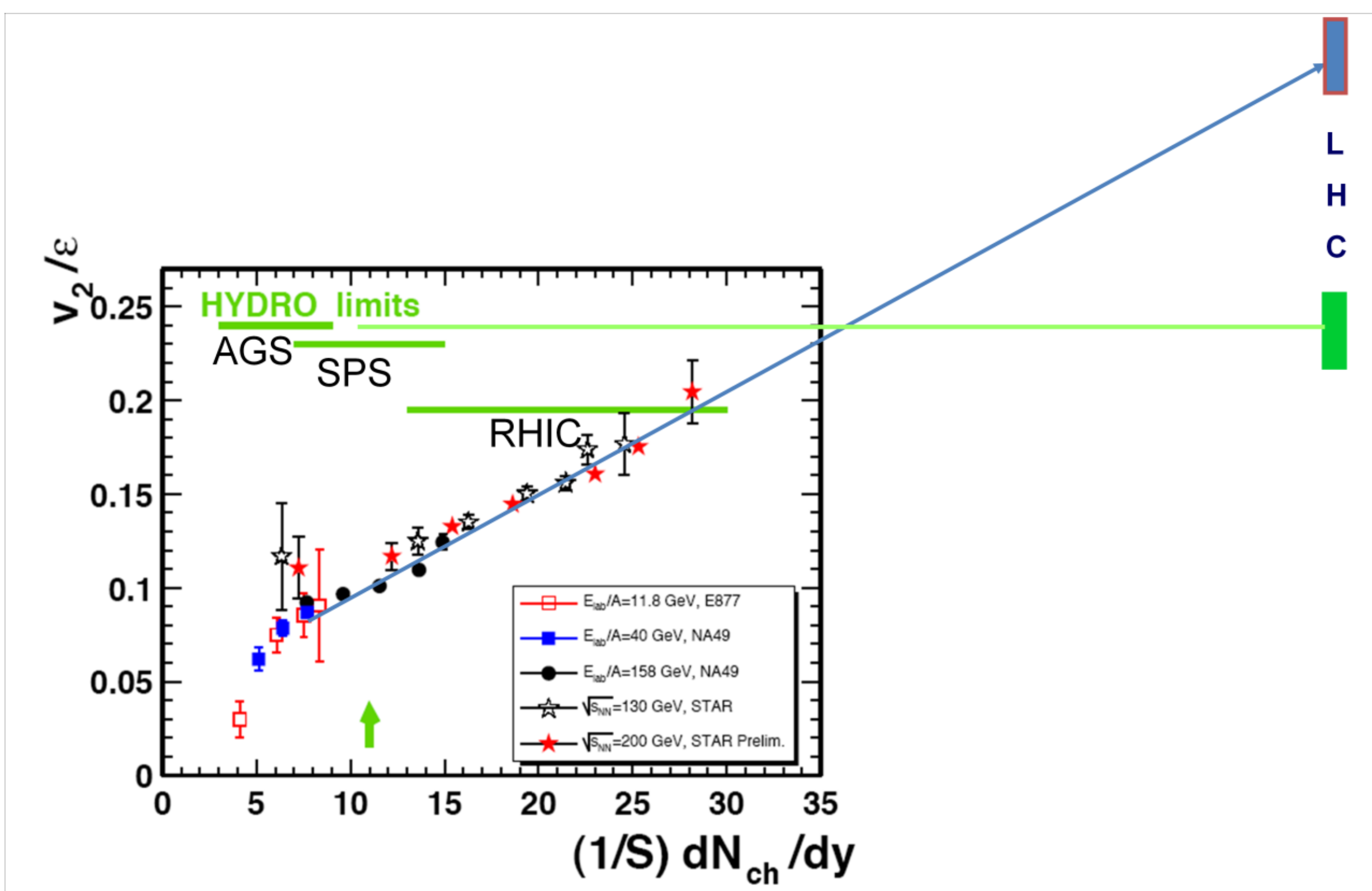

Figure 3: Elliptic flow divided by the initial eccentricity of the collision as a function of the multiplicity density, from [3].

Another early result which will bear major consequences on our understanding of high-density QDC matter will be elliptic flow, or $v_2$, the second Fourier coefficient of the azimuthal anisotropy of particle production. The RHIC data show a "perfect liquid" behavior, in agreement with hydrodynamic models. In figure 3, from [3], a compilation of elliptic flow data taken at different energies is shown. Hydrodynamics would predict a modest increase, while a simple extrapolation from the data would favour a much larger value. As in the case of multiplicity, a fairly simple measurement available within a short time from the first collisions will challenge our frame of understanding of the RHIC results!

Furthermore, thanks to the copious multiplicity of produced particles, a short low-luminosity run will allow general yet essential measurements such as the particle composition and the transverse momentum distributions of identified particles, providing the freeze-out temperature and the collective motion of the particle emitting source from the transverse momenta of particles and verifying the "hadrochemistry" thermal models which have successfully described hadron production up to RHIC energies [16]. ALICE, for example, will be able to reconstruct about 13 $\Lambda$, 0.1 $\Xi$ and 0.01 $\Omega$ particles per event and therefore collect a statistically relevant sample with as little as $10^6$ events. More generally, a sample of the order of $10^6$ events will give access to most of the measurements of the characteristics of the particle emitting source: particle spectra, resonances, differential flow analysis and interferometry. As an example among many, in Figure 4 are shown the invariant mass distributions, as measured in ALICE with $10^6$ central Pb-Pb events, for the $\rho$ going to $\pi^+\pi^-$ (left) and for the $\phi$ going to $K^+K^-$ (right). The mass resolution that ALICE can achieve even with a modest statistical sample is remarkable, respectively around 3 and 1 MeV.

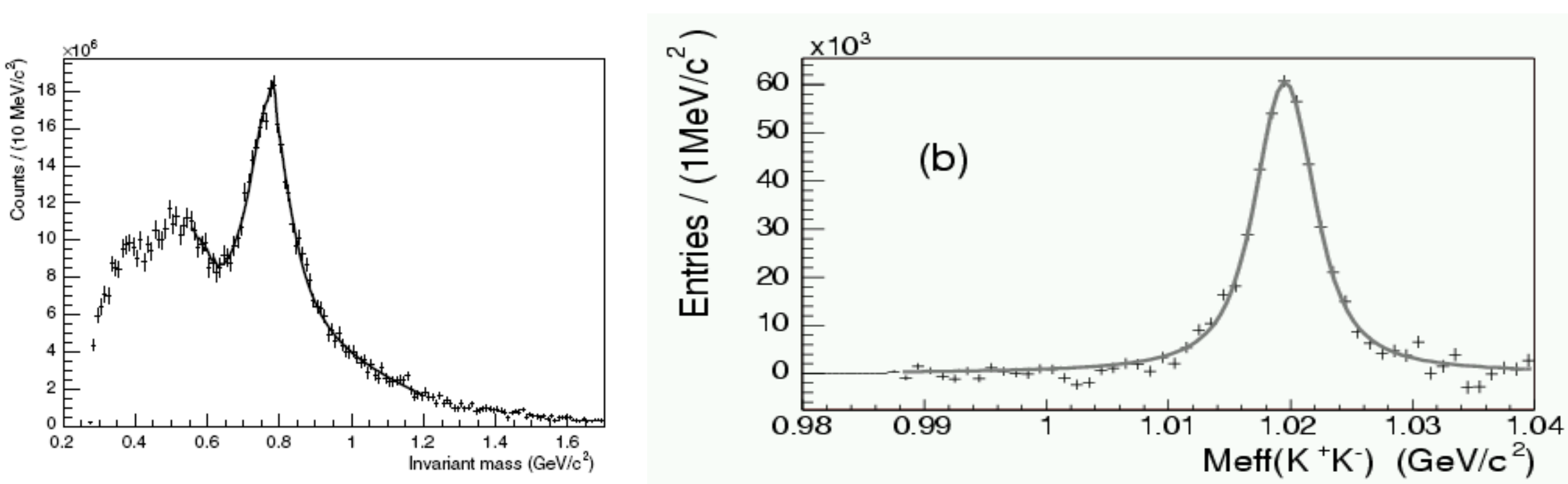

Figure 4: Examples of hadronic measurements in ALICE for $10^6$ central Pb-Pb events: invariant mass distributions for the $\rho$ going to $\pi^+\pi^-$ (left) and for the $\phi$ going to $K^+K^-$ (right), from [3].

## 5. LHC'S UNIQUE ASSETS

This chapter could read simply: "cross sections"! In Figure 5 are collected on a single plot the cross sections, and the rates at LHC Pb-Pb nominal luminosity, for several relevant "hard processes". The jump from RHIC to the LHC is huge, allowing a completely new, differential approach to the study of jets and heavy flavors. Cross section for heavy flavors grows by: $\sigma_{cc} \rightarrow \sim 10$ and $\sigma_{bb} \rightarrow \sim 100$, while the increase in $\sigma$ for the production of jets with transverse energy larger than 100 GeV is of several orders of magnitude. In particular, about $10^5$ jets with transverse momentum above 100 GeV/$c$ will be recorded by ALICE within the acceptance $|\eta| < 0.9$ during a nominal year of Pb beam at the LHC, and $1.5 \times 10^3$ in the $10^6$ events of a low-luminosity pilot run [18]. Thus, also to start the exploration of high-$p_T$ processes we will not have to wait for the full luminosity to come…

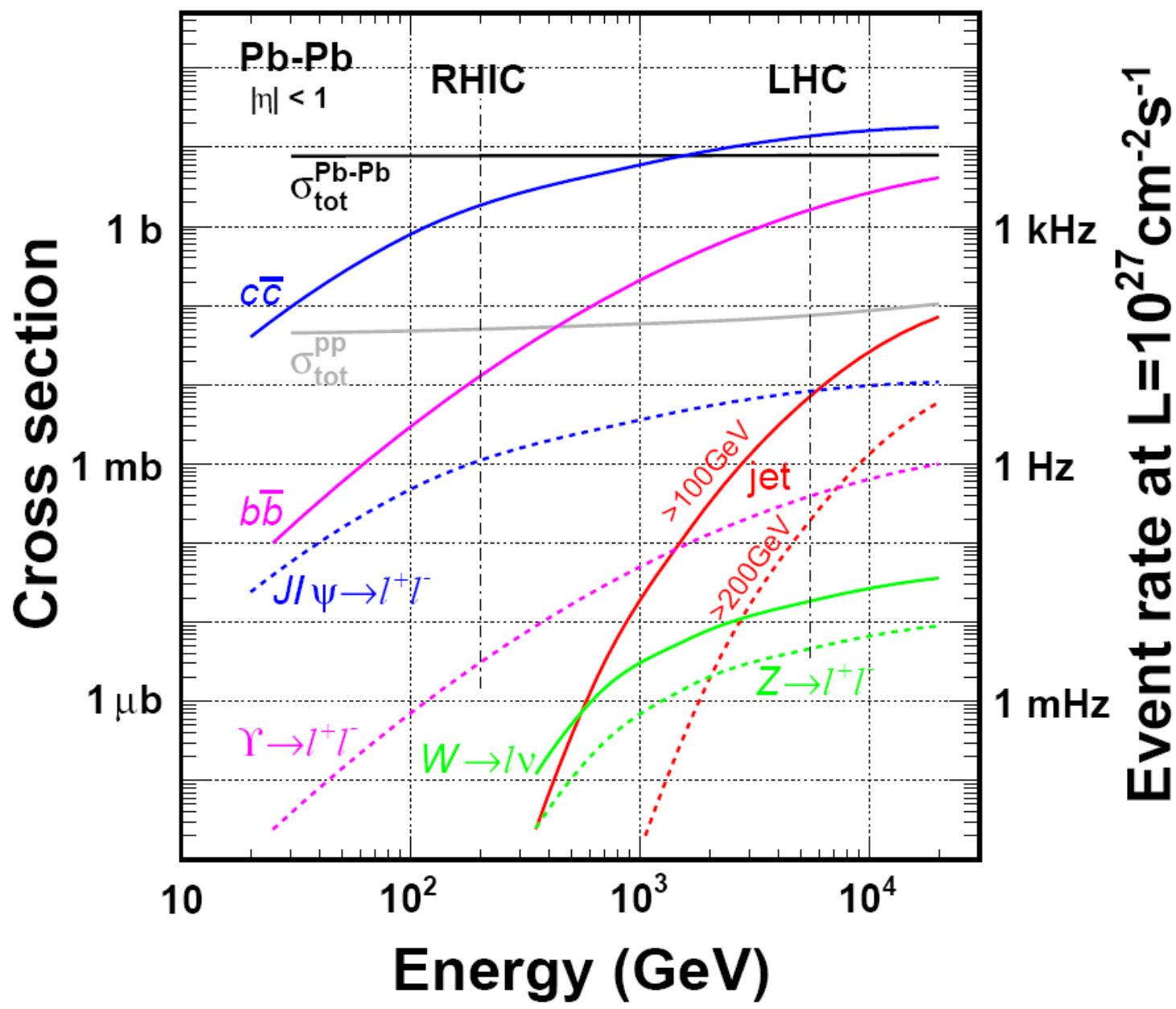

Figure 5: Cross sections (left axis) and event rates (right axis) for designated hard processes as a function of c.m. energy $\sqrt{s_{NN}}$, spanning RHIC and the LHC, from [17].

### 5.1. Jets

A hard parton traversing dense strongly-interacting matter looses part of its energy, mainly due to gluon bremsstrahlung [19]. This effect is called jet quenching and it results in a decreased yield of high-$p_t$ particles, as observed at RHIC. The amount of energy loss is very sensitive to the medium density and to the geometry of the collision. A detailed study of jet quenching, as it is possible when jet energy and fragmentation function are directly measured, would shed new light on the nature of the dense state created in the collision.

Jets produced by high-enough energetic partons are easy to detect even on top of the large soft-hadron background produced on heavy-ion collisions, as shown in Figure 6. Cone jet-finding algorithms have been tested, giving reconstruction efficiencies of about 80% for 50 GeV jets, rising to about 100% for jets with energy above 100 GeV. Both CMS and ATLAS, thanks to their excellent hadron and EM calorimeters, report a resolution on jet-energy around 15% for 100 GeV jets, while for ALICE, which uses the tracking system for charged particles combined with the electromagnetic calorimeter, a resolution of 34% for 100 GeV jets is predicted. Moreover, the triggering capability and the acceptance of the calorimeters will provide ATLAS and CMS with a much larger statistical sample, allowing for detailed studies also of di-jet events. Still, ALICE will provide complementary and more detailed information on the jets, thanks to the tracking of low-$p_t$ particles and the particle identification. In fact, the gluons radiated by an energetic parton inside the dense matter fragment into hadrons which mostly fall within the jet cone produced by the original parton [20]. Jet quenching therefore involves mostly a medium-induced redistribution of the jet energy inside the jet cone. It has been estimated that a 100 GeV jet loses about 20% of its energy but only about 3% of this energy falls outside the $R$ = 0.3 cone. To access the fragmentation of semi-hard and soft gluons inside the jet cone detailed jet characteristics, such as the hadron multiplicity, jet shape, fragmentation function, transverse momentum distribution with respect to jet axis will need to be measured and compared to those observed in pp interactions. These measurements require the track reconstruction and identification down to very low momenta, where the largest modifications are expected.

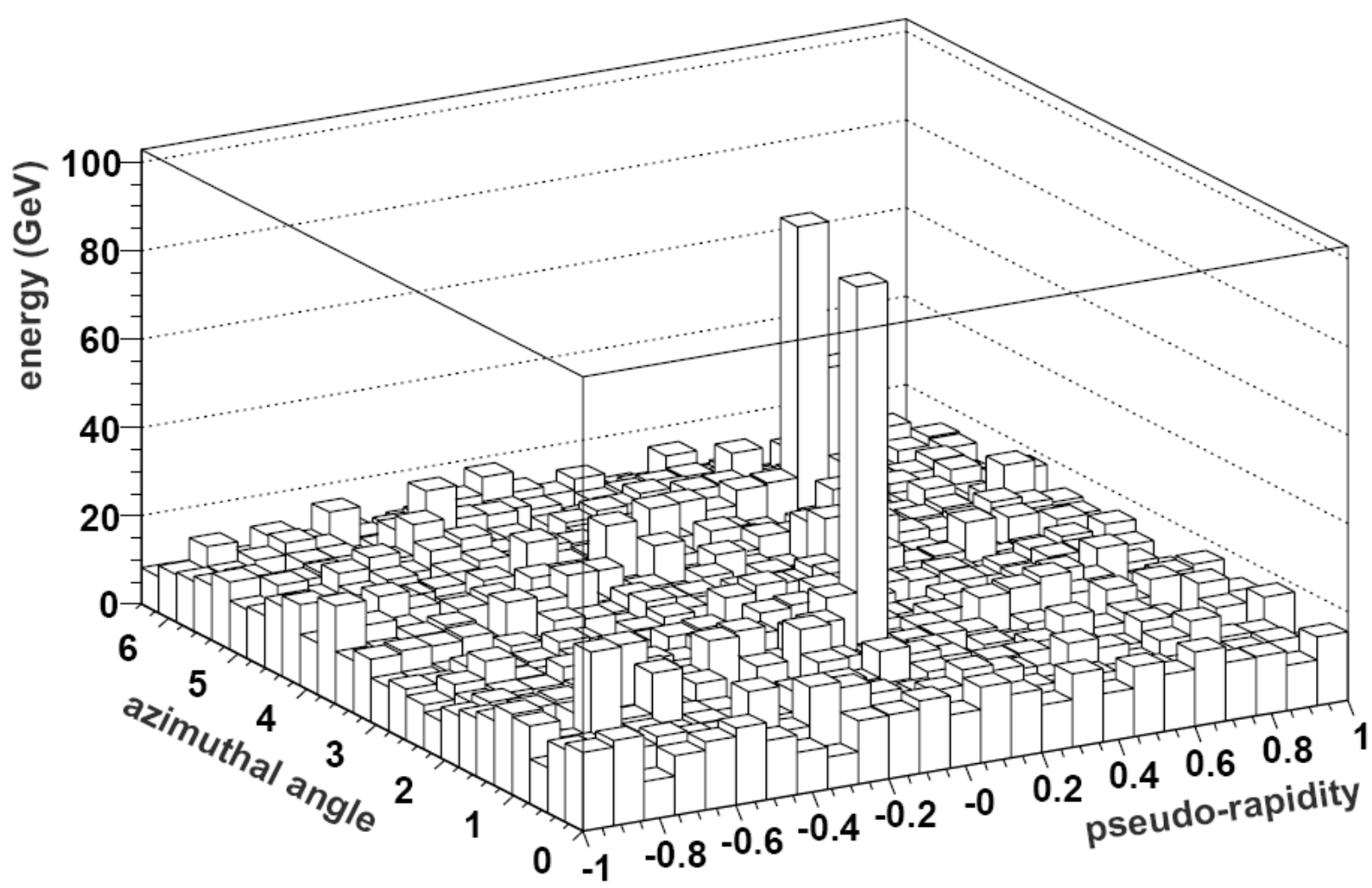

Figure 6: Pseudo-rapidity–azimuthal-angle plot of Pb–Pb event at LHC energy with two 100 GeV jets generated with HIJING and PYTHIA, from [18].

On the other hand, a precise measurement of the jet energy will be necessary to normalize the observables indicated above. Apart from instrumental resolution, there are irreducible contributions to the resolution in the measurement of the jet energy, such as the fluctuations in the energy of particles from the underlying event inside the jet cone. A more effective approach could be to use the energy of a recoiling particle opposite to the jet, in particular a weakly interacting one, such as a photon or a $Z_0$ boson. Jet–photon events will be produced abundantly, about $10^6$ events with photon energy above 50 GeV during a nominal heavy-ion run. ATLAS, thanks to the very large acceptance and unique photon identification capabilities of its calorimeters, will be the ideal detector for this type of studies. Tagging with $Z_0$ bosons would have the advantage of being practically background free, but the expected statistics is very limited, just few hundreds of events for a nominal heavy-ion run, so it will be challenging to apply for parton fragmentation function studies.

## 5.2. Heavy flavors

Since the historic 1986 prediction of J/y suppression in a plasma due to color screening [21], the study of the production of heavy flavors in high-energy nuclear collisions has been a centerpiece of this field of research. Heavy quarks are produced in hard parton scattering, during the initial stage of heavy-ion collisions, and therefore they provide a direct window to the early state of the system created in the collision. In addition, at least charm quarks may be produced in the QGP phase, if the temperature is high enough to overcome their production threshold. For a clear understanding of the processes involved, it is paramount to measure a variety of quarkonium states, with different radii and binding energies, over a wide range of transverse momenta. In fact, the larger the state, the sooner it is dissociated due to Debye screening in deconfined strongly-interacting matter, while high-*p*t quarkonia are less affected, since they spend a shorter time in the deconfined matter. The LHC is the ideal machine for heavy quarkonia studies, since the production rates are high enough for accurate measurements of the yields and $p_t$-spectra of most of the interesting states: J/ψ, ψ’, Y, Y*’* and Y*”*. ALICE and CMS will measure quarkonia with similar statistical significance, since ALICE will have lower background, while CMS will have better statistics and invariant mass resolution. Moreover, the two experiments provide complementary measurements, since CMS covers the central rapidity region, while ALICE extends the measurement to the forward region and, in the central region where quarkonia are measured in the dielectron channel, covers much lower transverse momenta. In the left panel of figure 7 is shown the invariant mass spectrum for dimuons in the Y region as expected to be measured by CMS. Given the uncertainties in the predicted production rate for heavy-quarks, it is important that quarkonia cross-sections are normalized to the production rates of another particles containing the same heavy flavor, namely D or B mesons. A precise measurement of open heavy-flavor production requires either direct meson detection down to low-*p*t, or a measurement of the dimuon-mass spectrum with very low pion- and kaon-decay background. Moreover, since at the LHC energy a large fraction of J/ψ are produced by the decay of B mesons, the direct measurement of J/ψ from B decays would be essential. ALICE does in fact feature all of the above, thus making the heavy flavor meson measurements on of the most promising items in its physics program, making full use of the combined tracking, vertexing and identification potential of the experiment.

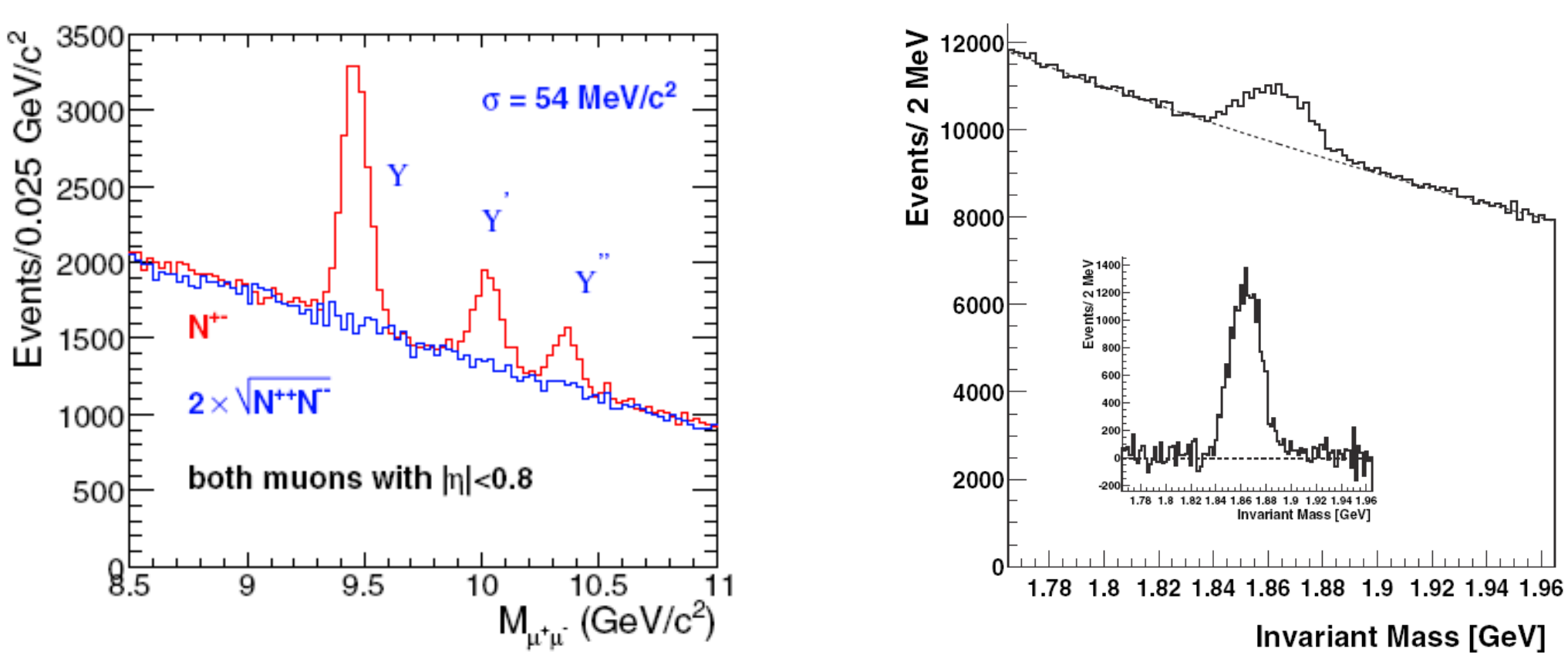

Figure 7: Left: Invariant mass spectra of opposite-sign and like-sign muon pairs as measured by CMS (from [4]) in the Y mass region. Right: $K^-\pi^+$ invariant mass distribution for $10^7$ events as measured by ALICE (from [3]). The same distribution after background subtraction is shown in the inset.

ALICE has the capability to measure several heavy-flavored particles identified in different decay channels [22]. The best studied one is at the moment the $D^0$, using the $D^0 \rightarrow K^-\pi^+$ decay mode [23]. In the right panel of figure 7 is shown the $K^{-}\pi^{+}$ invariant mass distribution for $10^7$ events as measured by ALICE. The $D^0$ can be reconstructed in a wide transverse momentum range, from 0.5 to 15 GeV/*c*, thus allowing a precise measurement of the total production cross-section. B mesons are measured inclusively via their semileptonic decay channels (B → e+X).

## 6. CONCLUSIONS

The LHC is a fantastic discovery machine, and its Heavy Ion program promises to reward the large and lively community preparing the experiments with a rich harvest of Physics results. The experiments will be commissioned with proton beams, so they will be ready to make full use of the nuclear beams as soon as they will be available. Already a short, low-luminosity pilot run would allow for fundamental new measurements, bringing new light into our understanding of strong interactions.